\documentclass[pra,aps]{revtex4}
\usepackage{graphicx}
\usepackage{amsmath}
\begin{document}
\title{Single impurity atom embedded in a dipolar two-soliton molecule as a qubit }
\author{S. M. Al-Marzoug$^{1}$, B. B. Baizakov$^{2}$, U. Al Khawaja$^3$ and H. Bahlouli$^{4}$}
\affiliation{$^{1}$ Physics Department and IRC-Advanced Quantum
Computing, King Fahd University of Petroleum and Minerals, Dhahran 31261, Saudi Arabia,\\
$^{2}$ S. A. Azimov Physical-Technical Institute of Uzbekistan Academy of Sciences, Tashkent, Uzbekistan,\\
$^{3}$ Department of Physics, School of Science, The University of
Jordan, Amman, 11942, Jordan. \\
$^{4}$ Physics Department and IRC-Advanced Materials, KFUPM, Dhahran
31261, Saudi-Arabia }
\date{\today}

\begin{abstract}
We consider a single impurity atom trapped in a double well (DW)
potential created by a dipolar two-soliton molecule in a
quasi-one-dimensional geometry. By solving the eigenvalue problem
for the impurity atom in the DW potential, we find that its ground
and first excited states are well separated from higher excited
states. This allows it to be approximated by a desirable two-level
quantum system. Numerical simulations of the Schr\"odinger equation,
governing impurity atom,  demonstrate periodic oscillations in the
probability of finding the impurity confined either to the ``left"
or to the ``right" well of the DW potential. An analytic expression
for the coherent oscillations of the population imbalance between
the two wells of the DW potential has been derived using the
two-mode approximation. Theoretical predictions of the mathematical
model are in good agreement with the results of numerical
simulations. Potential usage of the developed setup as a physical
realization of ``qubit" has been discussed.
\end{abstract}
\maketitle

\section{Introduction}

The enhanced computational power of quantum computers will be
primarily due to the use of quantum bits called {\it qubit}. Unlike
its classical counterpart, called {\it bit}, which enables
operations using only two digits ``0" and ``1", qubits can exist in
multiple states simultaneously, owing to the superposition principle
of quantum mechanics, thus leading to an enormous increase in speed
of operations. To date, the qubits have been realized on different
physical platforms, such as Josephson junctions, quantum dots,
nuclear magnetic resonances, trapped ions, neutral atoms in optical
tweezers, cold atoms in optical lattice, coherent photons, etc. (see
e.g. book \cite{nakahara-book}). Each of these realizations has
distinct advantages and limitations, and lack of a universally
accepted platform motivates the search for alternative realizations.
Impurity atoms confined to matter-wave bright and dark solitons in
the context of their qubit properties were investigated in Refs.
\cite{shaukat2017,shaukat2019}. The key benefit of this approach is
that the qubit is well protected from
decoherence~\cite{salerno2005}. Previous research primarily examined
single hump matter-wave solitons with an embedded impurity atom. The
energy levels of a particle in a single well potential, such as
Gaussian \cite{nandi2010} or P\"oschl-Teller
type~\cite{flugge-book,senn1986}, are well known to be considerably
spaced apart, making them less suitable for the functioning of a
qubit with a desired frequency and tunability. In contrast,
double-well potentials exhibit closely spaced energy levels, making
them suitable for qubit representation \cite{foot2011}. In this
context, tunnel-coupled semiconductor double quantum dots were
considered highly promising systems \cite{divincenzo2005,wiel2003}.

Here, we propose a qubit realization using a single impurity atom
embedded in a dipolar two-soliton molecule that forms a self-induced
double well (DW) potential. This configuration enables a robust
two-state quantum system, while the dipolar nonlocality offers
in-situ tunability of the barrier height/width and energy splitting
(qubit frequency) through the strength of inter-atomic interactions.
The present model illustrates a qubit that is delocalized in space,
with its computational basis states determined by the presence or
absence of an impurity atom in one of the two potential wells of a
double-well trap. A notable advantage of this approach is the
ability to create highly entangled states involving multiple qubits
\cite{mompart2003}. It is important to note that matter wave
solitons in Bose-Einstein condensates (BEC) with only contact atomic
interactions do not support stable bound states resembling
molecules. Therefore, ordinary BECs cannot provide a stable DW
potential for the impurity atom. In a different context, the
applications of BECs for quantum computation and information
processing have been reported in the literature
\cite{byrns2012,xu2022,boudjemaa2025,barshilia2024,shaukat2020,
shaukat2020a}.

Our strategy in this work goes as follows: we first develop a
variational approach to find the stationary state of the dipolar
two-soliton molecule, then numerically calculate the energy spectrum
of the impurity atom trapped in the molecular potential. In the
framework of the two-mode approximation we derive the frequency of
coherent oscillations of the population imbalance between the two
wells of the DW potential. To validate our findings, the analytical
predictions will be compared with numerical simulations. Finally, we
discuss the potential usage of our results in the design of a
soliton-based qubit.

\section{The wave profile of the two-soliton molecule}

The essential features of the present system can be understood by
considering the model in a quasi-one-dimensional geometry. Reducing
the coupled 3D Gross-Pitaevskii equations (GPE) to a quasi-1D form
was elaborated in numerous publications, among which we refer to a
recent paper devoted to binary dipolar BECs \cite{adhikari2024}. The
results of this work suggest that the influence of the minority
component (single non-dipolar atom) on the majority component
(soliton molecule consisting of ${\cal N} \sim 10^3 \div 10^4$
dipolar atoms) is proportional to $1/{\cal N}$. Consequently, the
effect of the impurity on the soliton molecule can be considered
negligible. In these conditions, the model is described by a 1D
nonlocal GPE supporting a two-soliton molecule and a linear
Schr\"odinger equation for a single impurity atom embedded in the
molecular potential
\begin{eqnarray}
i\psi_t &=& -\frac{1}{2}\psi_{xx} - q|\psi|^2 \psi - g \psi
\int^{\infty}_{-\infty}R(|x-x^{\prime}|)|\psi(x^{\prime},t)|^2 d
x^{\prime}, \label{gpe} \\
i\phi_t &=& -\frac{1}{2} \phi_{xx} - \gamma|\psi|^2 \phi, \label{se}
\end{eqnarray}
where $\psi(x,t)$ is the mean field wave function of the dipolar
soliton molecule, $\phi(x,t)$ is the wave function of the impurity
atom, $q$ and $g$ are the strengths of contact and dipole-dipole
attractive forces between atoms of the condensate. In experiments,
both of these parameters can be tuned, the former by a Feschbach
resonance method \cite{chin2010}, and the latter using rotating
magnetic fields \cite{tang2018}. In this setting, the impurity
``feels" the presence of a double well potential due to the dipolar
two-soliton molecule with a coupling strength $\gamma$.

The long-range interactions between atoms of the dipolar BEC is
characterized by a normalized kernel function, chosen to be of Gaussian type
\begin{equation}\label{kernel}
R(x)=\frac{1}{\sqrt{2\pi} w} \exp\left(-\frac{x^2}{2 w^2}\right),
\qquad \int_{-\infty}^{+\infty} R(x) dx = 1,
\end{equation}
where $w$ is a parameter that characterizes the nonlocal nature
of the dipolar atomic interactions in BEC. This expression is
selected for a computational convenience, although it is more
relevant to optical media. Qualitative similarity between the
Gaussian kernel function (\ref{kernel}) and other expressions used
in BEC research, such as the single mode approximation kernel
\cite{sinha2007} and that with a cutoff parameter \cite{cuevas2009},
justifies our choice.

The objective of this work is to develop an analytical framework for
the ``qubit problem", based on the variational approach (VA). The
validity of the VA for describing the properties of dipolar BEC over
a wide range of experimental parameters was demonstrated in
\cite{yi2001,olson2013}. It is evident from the basic model Eqs.
(\ref{gpe})-(\ref{se}) that the GPE is decoupled from the
Schr\"odinger equation, while the impurity ``feels" the dipolar BEC
only as an external potential. Obviously, a single impurity atom
cannot significantly distort the wave profile of the BEC, which
consists of a large number of dipolar atoms \cite{wenzel2018}.
Similar models have been employed, and proven to be realistic, in
previous publications devoted to BEC-based qubits
\cite{shaukat2017,shaukat2019,salerno2005}.

The wave profile of the two-soliton molecule in a dipolar BEC, that
is, the double well potential acting on the impurity, can be
determined using VA. To this end, we write the Lagrangian density
corresponding to GPE~(\ref{gpe})
\begin{equation}\label{lagden}
{\cal L} = \frac{i}{2}(\psi \psi^{\ast}_t - \psi^{\ast}\psi_t) +
\frac{1}{2} |\psi_x|^2 -\frac{q}{2}|\psi|^4 - \frac{g}{2} |\psi|^2
\int \limits_{-\infty}^{\infty} R(x-x^{\prime})
|\psi(x^{\prime},t)|^2 d x^{\prime}.
\end{equation}
A suitable trial function for the dipolar two-soliton molecule was
shown to have the form~\cite{baizakov2015}
\begin{equation}\label{ansatz}
\psi(x,t)=A(t) \, x \, \exp \left[ -\frac{x^2}{2a(t)^2} + i b(t) x^2
+ i\varphi (t) \right],
\end{equation}
where $A(t)$, $a(t)$, $b(t)$ and $\varphi(t)$ are variational
parameters, associated with the amplitude, width, chirp and phase,
respectively. The norm $N = \int |\psi(x)|^2 dx = A^2 a^3
\sqrt{\pi}/2$ is proportional to the number of atoms in the dipolar
condensate and it represents the conserved quantity of GPE
(\ref{gpe}). The main advantage of the trial function (\ref{ansatz})
over alternative two anti-phase Gussians \cite{otajonov2017} is that
it leads to simpler variational equations, although its validity is
limited to small deviations of solitons from their equilibrium
positions.

Substitution of the ansatz (\ref{ansatz}) and response function
(\ref{kernel}) into the Lagrangian density (\ref{lagden}) and
subsequent integration over the space variable $x$ yields the
effective Lagrangian
\begin{equation}\label{lagr}
\frac{L}{N} = \frac{3}{4a^2} + 3 a^2 b^2 + \frac{3}{2}a^2 b_t +
\varphi_t - \frac{3 q N}{8 \sqrt{2\pi} a} - \frac{g N }{8
\sqrt{2\pi}} \frac{3a^4+4a^2w^2+4w^4}{(a^2+w^2)^{5/2}}.
\end{equation}

\begin{figure}[htb]
\centerline{$\qquad \qquad (a)$ \hspace{7cm} $(b)$}
\centerline{
\includegraphics[width=8cm,height=6cm,clip]{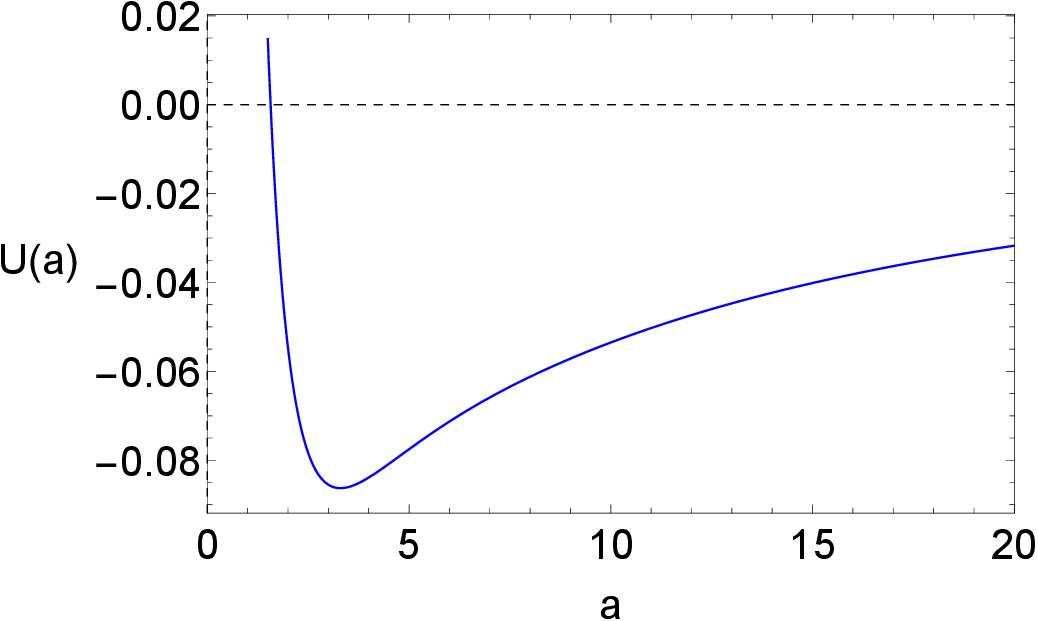}
\includegraphics[width=8cm,height=6cm,clip]{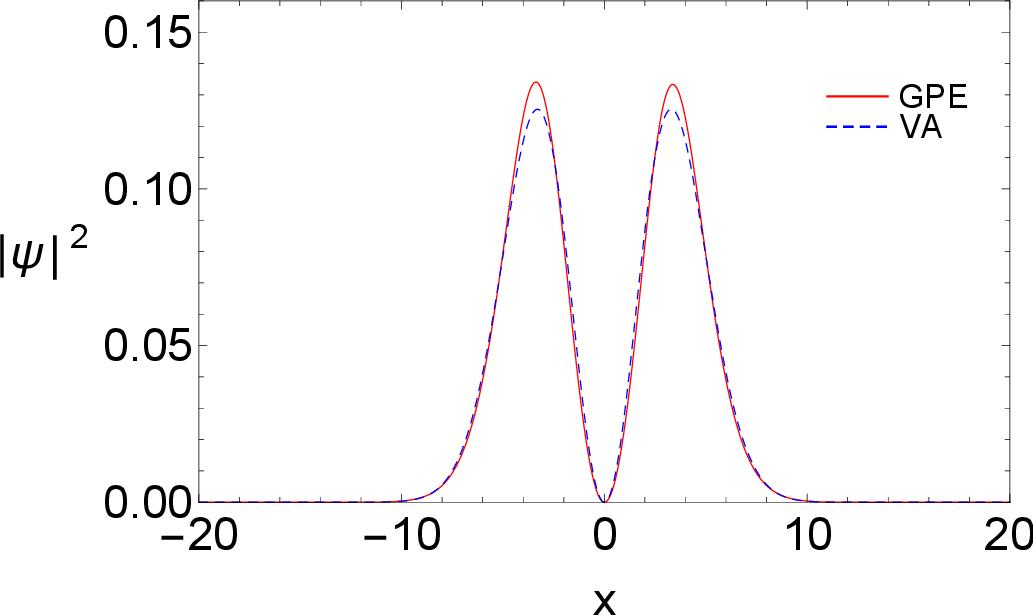}}
\caption{(a) The effective potential given by Eq. (\ref{pot}). (b)
The wave profile of the two-soliton molecule obtained by numerical
solution of the GPE (red line) and according to VA (blue dashed
line). Parameter values $A=0.177$, $a=3.299$ follow from VA
Eq.~(\ref{att}) for $q=1$, $g=6$, $w=5$, $N=1$. } \label{fig1}
\end{figure}

The VA equation for the parameter $a$ of the two-soliton molecule
can be derived from the corresponding Euler-Lagrange equations
$d/dt(\partial L/\partial \nu_t) - \partial L/\partial \nu = 0$ for
variational parameters $\nu \rightarrow a, b, \varphi$, using the
effective Lagrangian~(\ref{lagr})
\begin{equation}\label{att}
a_{tt}  = \frac{1}{a^3} - \frac{q N}{4\sqrt{2\pi} a^2} - \frac{g
N}{4 \sqrt{2 \pi}} \frac{a (a^4+4w^4)}{(a^2+w^2)^{7/2}}.
\end{equation}
This equation is similar to the equation of motion of a unit mass
particle in the anharmonic potential
\begin{equation}\label{pot}
a_{tt}=-\partial U/\partial a \quad \mbox{with} \quad U(a) =
\frac{1}{2 a^2} - \frac{q N}{4\sqrt{2\pi}a} - \frac{g N}{12
\sqrt{2\pi}} \frac{3a^4+4a^2w^2+4w^4}{(a^2+w^2)^{5/2}},
\end{equation}
which is plotted in Fig.~\ref{fig1}a for particular parameter
settings. The minimum of the potential corresponds to the stationary
width of the soliton molecule $a_0$, whose value can be found from
the fixed point of Eq. (\ref{att}), defined by $a_{tt} =  0$, for
particular parameter values $q$, $g$, $N$ and $w$. The separation
between center-of-mass positions of two anti-phase bright solitons
constituting the molecule is $\xi = 4 a_0/\sqrt{\pi}$. At larger
separation ($a>a_0$) the solitons attract each other ($\partial
U/\partial a > 0$), and at smaller separation ($a < a_0$) they repel
($\partial U/\partial a < 0$), therefore the effective potential
$U(a)$ has a binding nature necessary for the formation of the
stable molecule \cite{baizakov2019}. The wave profile of the dipolar
two-soliton molecule, numerically constructed using the imaginary
time propagation method \cite{chiofalo2000} applied to GPE
(\ref{gpe}), and as predicted by VA, is shown in Fig. \ref{fig1}b.

The stability of the soliton molecule has been verified by real-time
GPE evolution of the spatially perturbed initial wave profile.
During time propagation, the soliton molecule sheds the perturbation
as linear waves, which are absorbed at the integration domain
boundaries and quickly acquires a smooth profile, thereby
demonstrating its stability.

\section{Bound states of the impurity atom trapped by a two-soliton molecule}

Once the wave profile of the two-soliton molecule has been
determined, the bound states of the impurity atom subject to the
molecular potential can be identified from the eigenvalue problem associated
with Eq.~(\ref{se})
\begin{equation}\label{eig}
H \phi_j(x) = \mu_j \phi_j, \qquad j=0,1,2,...,
\end{equation}
where $H = -\frac{1}{2}\partial_x^2 + V(x)$ is the Hamiltonian, and
$V(x) = -\gamma |\psi(x)|^2$ represents the double-well potential
created by the dipolar two-soliton molecule. For numerical
implementation, we discretize the spatial domain $x \in [-L/2, L/2]$
and utilize the matrix representation of Eq. (1). Then, using the
standard procedure {\tt Eigensystem[H]} of the Mathematica package,
we obtain bound state energies and corresponding wave functions of
the impurity atom in the DW potential.

\begin{figure}[htb]
\centerline{$\qquad \qquad (a)$ \hspace{7cm} $(b)$}
\centerline{
\includegraphics[width=8cm,height=6cm,clip]{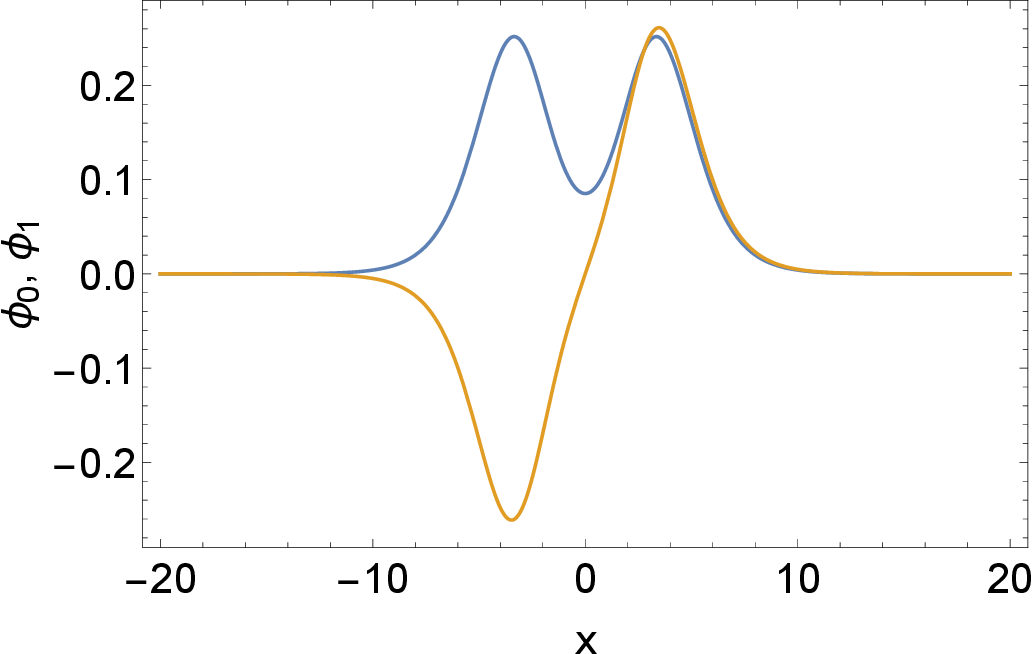}
\includegraphics[width=8cm,height=6cm,clip]{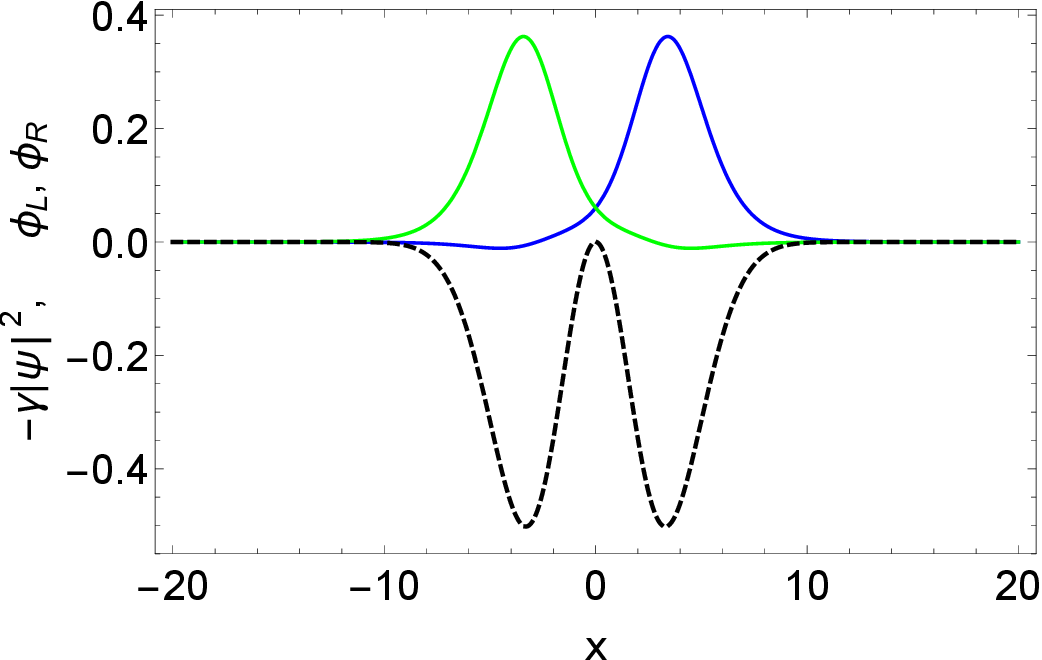}}
\caption{(a) The wave profiles of the symmetric ground state with
$\mu_0 = -0.326$ (blue), and the antisymmetric first excited state
with $\mu_1=-0.312$ (yellow) of the impurity atom in the double well
potential. (b) Localized combinations constructed from the linear
superposition of the ground and first excited states of the system
according to Eq. (\ref{lr}). For visual convenience the double well
potential $V(x)=-\gamma |\psi(x)|^2$ with $\gamma=4$ is also shown
by a dashed black line. The parameter values are similar
to~Fig.~\ref{fig1}.} \label{fig2}
\end{figure}

The ground state ($\mu_0 = -0.326$) and first excited state
($\mu_1=-0.312$) wave profiles of the system~(\ref{eig}) are shown
in Fig. \ref{fig2}a. Evidently, the energy levels of the symmetric
ground state $\phi_0(-x) = \phi_0(x)$, and antisymmetric first
excited state $\phi_1(-x) = -\phi_1(x)$ are close. In addition the
doublet components ($\phi_0,\phi_1$) are real and orthonormal with
energy splitting $\Delta = \mu_1-\mu_0 >0$
\begin{equation}\label{ort}
\int_{-\infty}^{\infty}\phi_j(x) \phi_k (x)dx = \delta_{j,k}, \quad
j,k = 0,1.
\end{equation}
The next excited states with a negative energy $\mu_2=-0.060$ and
$\mu_3=-0.018$ are well separated from the lowest two states. This
indicates that the dipolar two-soliton molecule with an embedded
impurity atom can be approximated as a two-level system. Since any
two-level quantum system can function as a qubit, thus the proposed
model is appropriate for our purpose.

To observe coherent Josephson type oscillations of the impurity atom
in the molecular double well potential we construct the localized
combinations (see Fig. \ref{fig2}b)
\begin{equation}\label{lr}
\phi_R(x) = \frac{\phi_0(x)+\phi_1(x)}{\sqrt{2}}, \qquad  \phi_L(x)
= \frac{\phi_0(x)-\phi_1(x)}{\sqrt{2}},
\end{equation}
whose orthonormality can easily be verified. Physically, $\phi_R$
($\phi_L$) are confined in the right (left) well of the DW potential
because the symmetric ($\phi_0$) and antisymmetric ($\phi_1$) waves
localize when combined as indicated in Eq. (\ref{lr}).

When the initial state for Eq. (2) is selected as corresponding to
$\phi_R$ or $\phi_L$ in Eq. (\ref{lr}), the two-level system evolves
smoothly in time such that the probability of finding the atom in
the left or right well of the DW potential periodically changes. The
frequency of these oscillations is known as a Rabi frequency
\cite{griffiths-book}. External driving fields at Rabi frequency are
used for calibration and manipulation of qubits. It is essential to
note that the system always remains in a superposition of the two
basis states \{$\phi_0, \phi_1$\}, whose probability amplitudes
change continuously unless the decoherence mechanisms, including the
attempt to measure the system's state, come into play.

\section{Dynamics of the impurity atom in a double-well potential}

The time evolution of the system can be described by the celebrated
two-mode model \cite{smerzi1997}. To this end, we approximate the
full wavefunction by projecting it onto the subspace spanned by the
lowest energy modes $\{\phi_L,\phi_R\}$
\begin{equation}\label{ansatz1}
\phi(x,t) \approx c_L(t)\,\phi_L(x) + c_R(t)\,\phi_R(x),
\end{equation}
where $c_L(t),c_R(t)$ are complex time-dependent amplitudes. The
normalization condition is given by
\begin{equation}\label{nrm}
\int |\phi(x,t)|^2 dx \approx |c_L|^2 + |c_R|^2 = 1.
\end{equation}
In the next step we derive the evolution equations for the amplitudes by
projecting the full equation (\ref{se}) onto the $\phi_L$ and
$\phi_R$ modes using the ansatz (\ref{ansatz1})
\begin{equation}
i\,\partial_t\big(c_L\phi_L + c_R\phi_R\big) = H \big(c_L\phi_L +
c_R\phi_R\big).
\end{equation}
Multiplying on the left by $\phi_L(x)$ and integrating over $x$ we
obtain an equation for $c_L$ (using orthonormality). The equation
for $c_R$ is obtained in a similar way. The result is a pair of
coupled linear ordinary differential equations~(ODE)
\begin{eqnarray}
i\,\dot{c}_L &=& \langle \phi_L | H | \phi_L \rangle c_L + \langle
\phi_L | H | \phi_R \rangle c_R ,
\label{cl}\\
i\,\dot{c}_R &=& \langle \phi_R | H | \phi_L \rangle c_L + \langle
\phi_R | H | \phi_R \rangle c_R. \label{cr}
\end{eqnarray}
The matrix elements are calculated as follows
\begin{equation}
\langle \phi_L | H | \phi_L \rangle = \langle \phi_R | H | \phi_R
\rangle = \frac{\mu_0+\mu_1}{2}, \quad \langle \phi_L | H | \phi_R
\rangle =\langle \phi_R | H | \phi_L \rangle = \frac{\mu_0-\mu_1}{2}
\;=\; \frac{\Delta}{2}.
\end{equation}
Thus, the matrix representation of $H$ in the R/L basis is given by

$$
H^{(LR)}= \left(
\begin{array}{cc}
(\mu_0+\mu_1)/2 & \Delta/2        \\
\Delta/2       & (\mu_0+\mu_1)/2  \\
\end{array}
\right).
$$
The diagonal elements $(\mu_0+\mu_1)/2$ only contribute to a global
phase factor and can be ignored since they cancel in the calculation of
probabilities and expectation values of observable quantities. Now,
defining the tunneling amplitude $J = \Delta/2 >0$, we can rewrite
the Eqs. (\ref{cl})-(\ref{cr}) as follows
$$
i \frac{d}{dt} \left(
\begin{array}{c}
c_L  \\
c_R
\end{array}
\right) = \left(
\begin{array}{cc}
0 & J  \\
J & 0
\end{array}
\right) \left(
\begin{array}{c}
c_L  \\
c_R
\end{array}
\right).
$$
This yields the pair of coupled first-order ODEs
\begin{equation}\label{odes}
i\,\dot{c}_L(t) = J\,c_R(t), \qquad i\,\dot{c}_R(t) = J\,c_L(t),
\end{equation}
where dots denote time derivatives. The final second-order ODE for
$c_L$ is
\begin{equation}\label{cL}
\ddot{c}_L(t) + J^2\, c_L(t) = 0.
\end{equation}
This is a simple harmonic oscillator ODE with a frequency $\Omega =
J$. A similar equation governs the other component $c_R(t)$. The
general solution of Eq. (\ref{cL}) is
\[
c_L(t) = A\cos\!\left(\Omega t\right) + B\sin\!\left(\Omega
t\right),
\]
with complex constants $A,B$ to be fixed by initial conditions. The
probabilities to find the atom in the left and right wells of
the DW potential are $P_L(t) = |c_L(t)|^2$ and $P_R(t) =
|c_R(t)|^2$, respectively. Obviously, the normalization condition
$P_L(t)+P_R(t) = 1$ holds.

In analogy with Josephson oscillations we define the population
imbalance $z(t) = P_L(t) - P_R(t) = 1- 2 P_R(t)$. Using the special
initial condition $c_L^{(0)}=1,c_R^{(0)}=0$ we obtain
\begin{equation}\label{popimb}
z(t) = \cos(2 \Omega t), \qquad \mbox{with} \qquad \Omega =
\Delta/2.
\end{equation}
Figure 3 illustrates the real-time dynamics of the coupled
system~(1)--(2) for the parameter set $q = 1$, $g = 10$, $\gamma =
3$, and $N = 10$. Panel~(a) shows the stationary double-well
potential $|\psi(x,t)|^{2}$ created by the dipolar two-soliton
molecule. As expected, the molecular density remains practically
unchanged during the evolution, confirming that the impurity atom
has a negligible back action on the condensate. Panel~(b) presents
the spatiotemporal evolution of the density $|\phi(x,t)|^{2}$, where
the impurity periodically tunnels between the left and right wells
of the self-induced potential, maintaining its localization and
phase coherence over long time scales. Panel~(c) displays the
corresponding population imbalance
\begin{equation}
z(t) = \int_{x<0} |\phi(x,t)|^{2}\,dx - \int_{x>0} |\phi(x,t)|^{2}\,dx,
\end{equation}
which demonstrates clear Josephson-like oscillations with a
well-defined period. The measured oscillation period from direct
numerical integration, $T_{\mathrm{meas}} = 223.506$, agrees
remarkably well with the theoretical prediction obtained from the
two-mode model, $T_{\mathrm{pred}} = 223.125$, in accordance with
Eq.~(20). This excellent agreement validates the analytical
approximation and confirms that the impurity atom behaves as a
coherent two-level system exhibiting Rabi-type oscillations between
the two potential wells.
\begin{figure}[htb]
\centerline{
\includegraphics[width=16cm,height=10cm,clip]{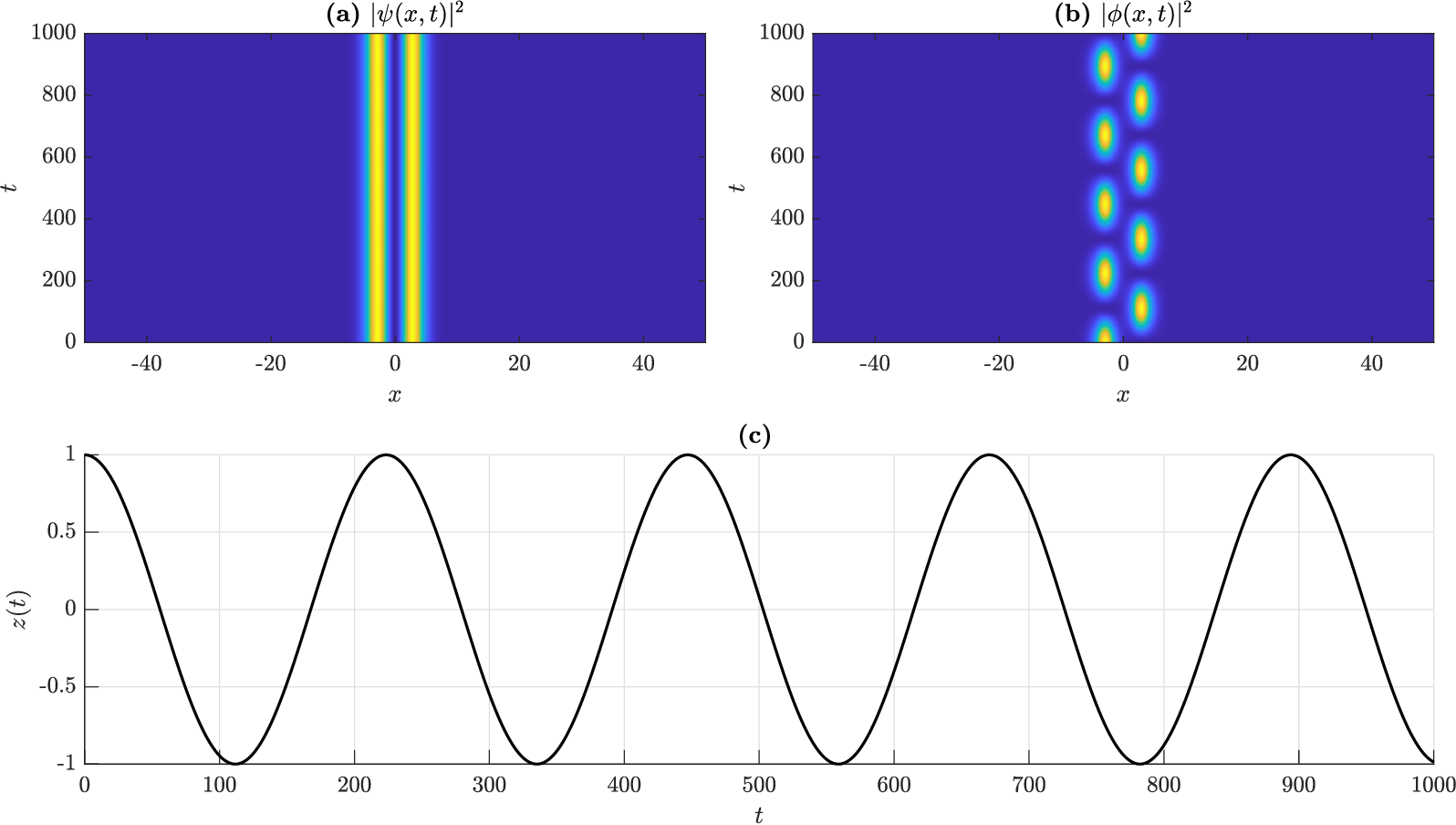}}
\caption{(a) The DW potential $\sim |\psi(x,t)|^2$ created by the
dipolar two-soliton molecule remains stationary as time advances.
(b) The density plot for $|\phi(x,t)|^2$ shows regular oscillations
of the probability of finding the impurity atom on the left/right
well. (c) The corresponding population imbalance
$z(t)=\int_{x<0}|\phi(x,t)|^{2}dx - \int_{x>0}|\phi(x,t)|^{2}dx$
demonstrates coherent Josephson-type oscillations. The numerically
measured oscillation period $T_{\mathrm{meas}} = 223.506$ agrees
closely with the theoretical prediction $T_{\mathrm{pred}} =
223.125$ obtained from the two-mode model [Eq.~(20)]. The parameters
used are $q = 1$, $g = 10$, $w=5$, $\gamma = 3$, and $N = 10$. }
\label{fig3}
\end{figure}

\section{Tuning the qubit's frequency}

\begin{figure}[htb]
\centerline{$\qquad \qquad (a)$ \hspace{7cm} $(b)$}
\centerline{
\includegraphics[width=8cm,height=6cm,clip]{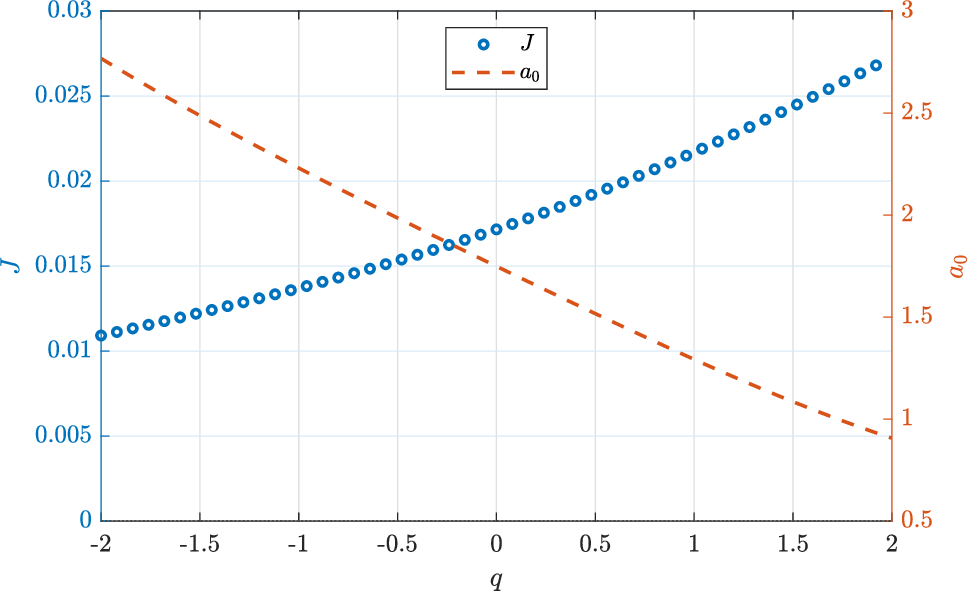}\qquad
\includegraphics[width=8cm,height=6cm,clip]{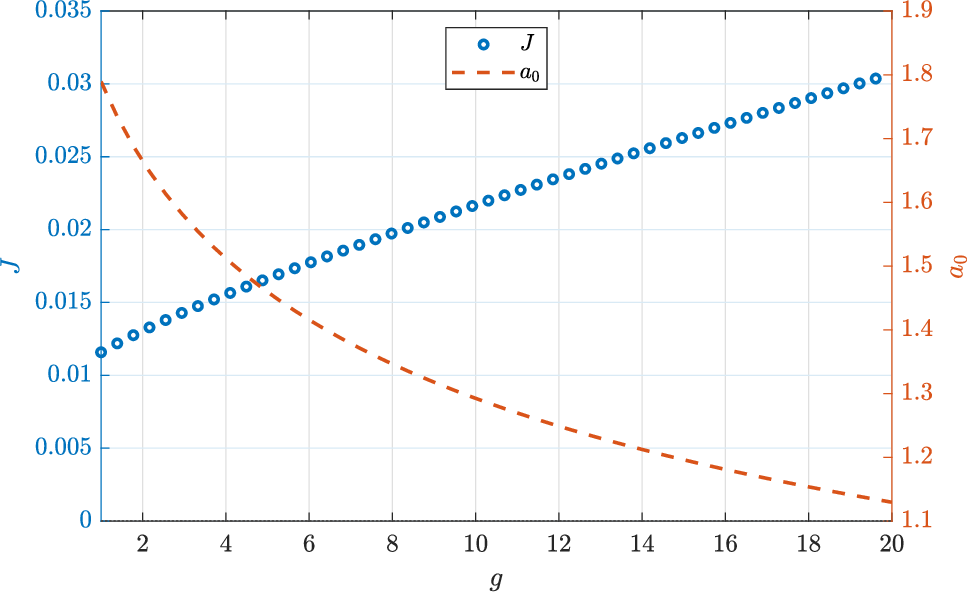}}
\caption{The frequency of coherent oscillations of the impurity atom
between the two wells of the double-well potential as a function of
the interaction strengths in the dipolar condensate. (a) Dependence
of the oscillation frequency on the contact interaction parameter
$q$ for fixed $g = 10$, $w=5$, $\gamma = 3$, and $N = 10$. (b)
Dependence on the dipolar (nonlocal) interaction strength $g$ for
fixed $q = 1$, $w=5$, $\gamma = 3$, and $N = 5$. Blue circles
represent the numerically obtained oscillation frequencies, while
the yellow dashed curves correspond to the analytical predictions
based on the variationally obtained width parameter $a_{0}$ from
Eq.~(8).} \label{fig4}
\end{figure}

The ability to change the qubit's frequency is crucial for its
precise control and manipulation. Since the qubit's frequency is
determined by the energy difference  $\Delta$ between its two lowest
quantum states $\{\phi_0, \phi_1\}$, the goal can be achieved by
manipulating the underlying DW potential $V(x) \sim -|\psi(x)|^2$ in
which the impurity atom is trapped. The DW potential can be modified
by tuning the strength of atomic interactions in BEC, either
short-range contact interactions $q$, or long-range dipolar
interactions $g$. The variation of these parameters will result in
corresponding variations of the distance $\sim a_0$ between the two
wells of the DW potential, thus the desired energy splitting $\Delta
= \mu_1 - \mu_0$. When the two potential wells are sufficiently
distant, the energy splitting disappears as the impurity atom's
tunneling probability approaches zero. In Fig. \ref{fig4} we show
the qubit's frequency as a function of the strength of contact and
dipolar atomic interactions. The curves demonstrate the tendency for
the energy splitting $J=(\mu_1-\mu_0)/2$ (proportional to the
qubit's oscillation frequency) to increase, as the two wells of the
DW potential approach each other (width parameter $a_0$ decrease).
Evidently, the effect of contact interactions is almost linear,
while the long range dipolar interactions show hyperbolic behavior.
The system's tunability facilitates controlled interactions between
neighboring qubits in an array, enabling the generation of entangled
states that are essential for robust quantum computations and
effective error correction schemes.

The dynamics of a qubit are best represented using the Bloch sphere.
The qubit, initialized in the $\phi_0$ state, precesses around the
XY-plane of the Bloch sphere. To create superposition states like
$(\phi_0 + \phi_1)/\sqrt{2}$, it is necessary first to initialize
the qubit in the state $\phi_0$, and then apply a Hadamard gate.

Figure 5 shows the Bloch sphere, which is generated directly from
numerical simulations. We first construct the DW potential due to
the dipolar two-soliton molecule, using the imaginary time
propagation method applied to the governing GPE (\ref{gpe}). Then,
using the obtained DW potential, we solve the eigenvalue problem for
Eq. (\ref{se}) to find the quantum states of the impurity atom
$\{\phi_0(x), \phi_1(x) \} $. Next, inserting these as initial
conditions,  we solve the coupled system (\ref{gpe})-(\ref{se}) in
real time. At each recorded time we project the impurity state onto
the two basis modes, thus finding a pair of complex amplitudes.
After re-normalizing, to ensure an exact two-level system, we
converted those amplitudes onto the Bloch coordinates $r_x = 2\,
{\rm Re} (c_0 c_1^{\ast}), \quad r_y = 2\, {\rm Im} (c_0
c_1^{\ast}), \quad r_z = |c_0|^2 - |c_1|^2$. That gives one point on
the sphere per time sample, forming the trajectory. Then, we plotted
the light sphere, drew the trajectory, and marked the start/end
points.
\begin{figure}[htb]
\centerline{
\includegraphics[width=6cm,height=6cm,clip]{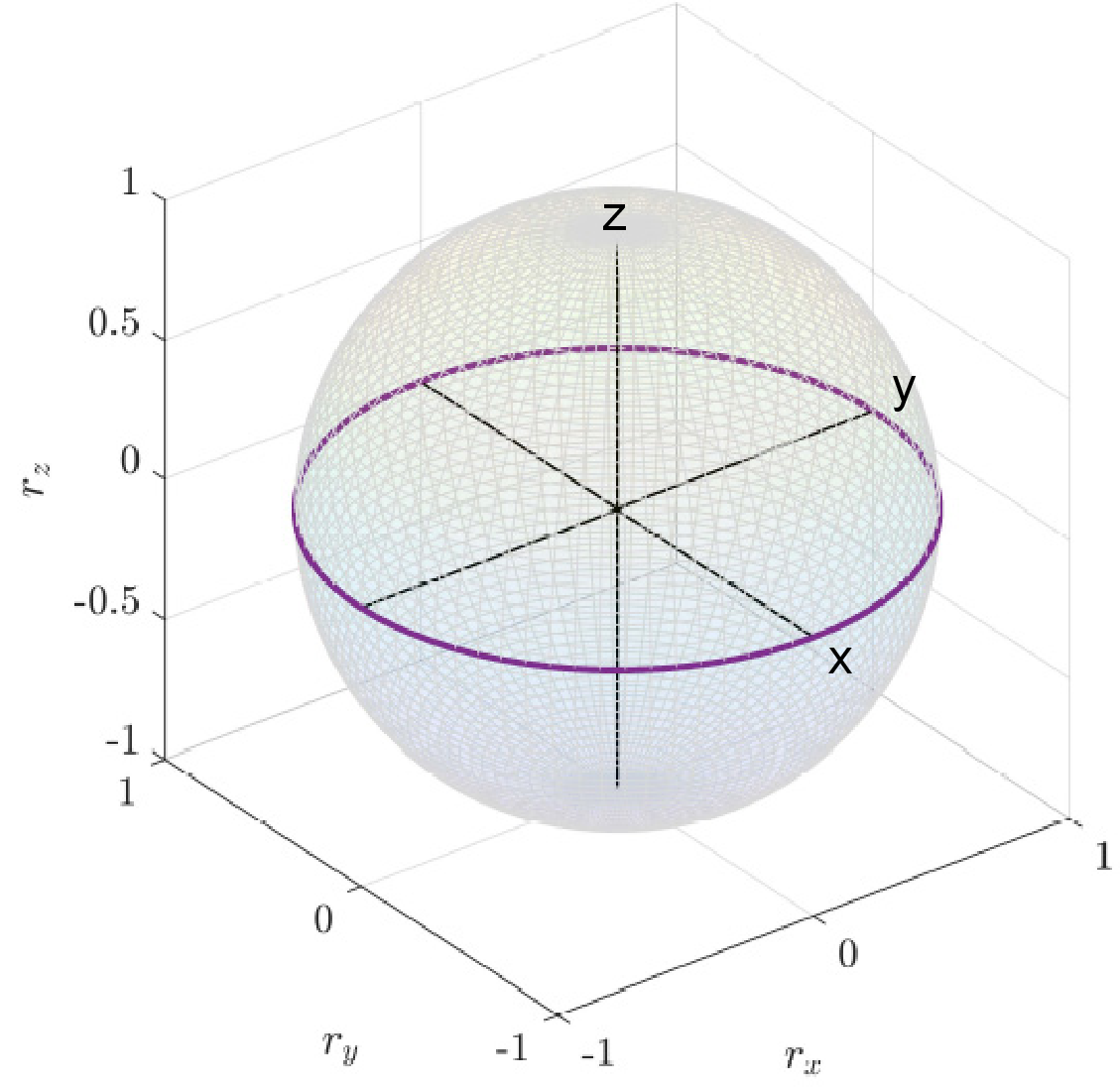}\qquad
\includegraphics[width=10cm,height=6cm,clip]{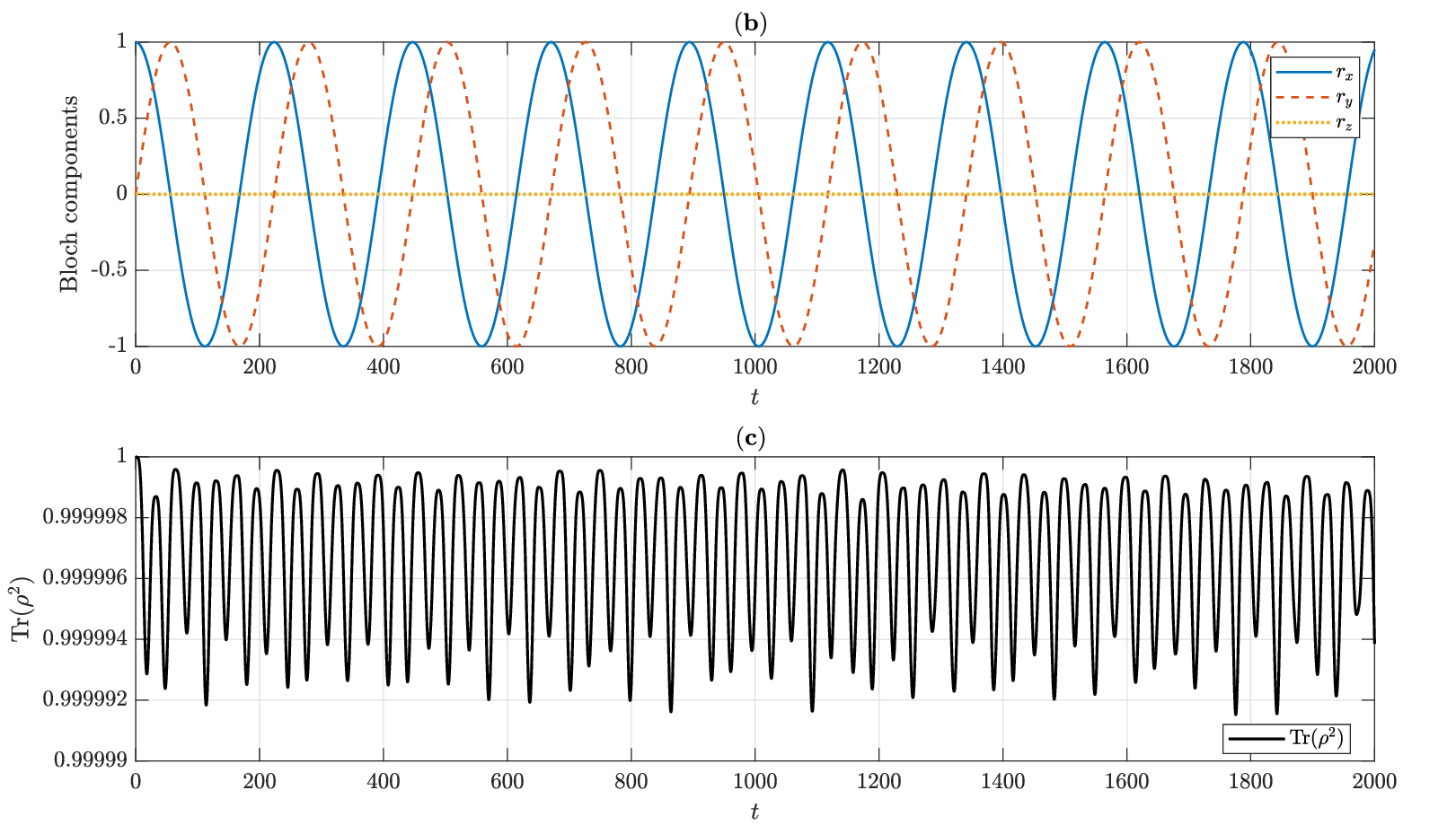}}
\caption{Left panel: The Bloch sphere illustrates the qubit state
generated from direct numerical simulations as described in the main
text. The north pole corresponds to the ground state $\phi_0$ and
the south pole corresponds to the first excited state $\phi_1$. An
arbitrary superposition of these basis states can be obtained by
parametrization $\phi = \cos(\theta/2)|\phi_0\rangle + e^{i \varphi}
\sin(\theta/2)|\phi_1\rangle$. Right panel: The Bloch components
$r_x \simeq \cos(\Delta t)$, $r_y \simeq \sin(\Delta t)$, $r_z
\simeq 0$ show circular equatorial motion (upper). The purity of the
two-level system calculated as ${\rm Tr}(\rho^2)$, with $\rho$ being
the density matrix, remains very close to unity indicating
negligible de-coherence (lower). The parameters used are $q = 1$, $g
= 10$, $w=5$, $\gamma = 3$, and $N = 10$.} \label{fig5}
\end{figure}

\section{Prospects of experimental realization}

The concept of this work is based on the properties of a two-state
quantum system formed by an impurity atom within a double-well
potential, induced by a dipolar two-soliton molecule. Although the
feasibility of creating soliton molecules in dipolar BECs has been
theoretically demonstrated in several publications
\cite{lakomy2012,edmonds2017}, their experimental realization has
not yet been reported. Meanwhile, soliton molecules, consisting of
stable bound states of two or more bright solitons, have been
experimentally observed in dispersion-managed optical fibers
\cite{stratmann2005,rohrmann2013} and mode-locked fiber lasers
\cite{tang2001,liu2022}. Advances in quantum technologies, based on
a soliton-impurity system, may inspire experiments to create dipolar
soliton molecules.

Numerical simulations presented in this work use the parameters of
the $^{164}$Dy condensate \cite{lu2011}, which has atoms with the
largest permanent magnetic dipole moment of $\mu = 10 \mu_B$, and
$s$-wave scattering length $a_s = 92 a_B = 4.87 \times 10^{-9}$m.
Strong radial confinement with a frequency of $\omega_{\bot} = 2\pi
\times 60$ Hz gives the system a quasi-one-dimensional nature. The
corresponding radial harmonic oscillator length is given by
$l_{\bot} = \sqrt{\hbar/(m \omega_{\bot})} \approx 1 \, \mu $m. The
characteristic length of dipolar interactions is estimated using the
formula $a_{dd} = \mu_0 \mu^2 m/(12 \pi \hbar^2) \simeq 7 \times 10
^{-9}$m, where $\mu_0 = 12.57$ N/A$^2$ is the permeability of
vacuum, $\mu = 9.274 \times 10^{-23}$ A m$^2$ and $m= 2.72 \times
10^{-25}$ kg are the magnetic moment and mass of the $^{164}$Dy
atom, respectively. The ratio of the dipolar to contact interaction
lengths, $\epsilon = a_{dd}/a_s \simeq 1.43$, indicates that we are
in the regime dominated by dipolar interactions. For this set of
parameters, the two-soliton molecule with dimensionless norm $N = 1$
and waist of the response function $w=5$, will contain $\sim 2.5
\times 10^{3}$ atoms. This number is compatible with the total
number of condensed atoms $\sim 15 \times 10^3$ produced in the
experiment \cite{lu2011}. Estimates for the other two dipolar
condensates, $^{168}$Er ($d=7 \mu_B, a_s = 68 \, a_B$)
\cite{aikawa2012,patscheider2022} and $^{52}$Cr ($d=6 \mu_B, \, a_s
= 170 \, a_B$) \cite{griesmaier2005,schmidt2003}, where the dipole
moment and $s$-wave scattering length are given in units of Bohr
magneton ($\mu_B$) and Bohr radius ($a_B$), also yield
characteristic parameters that are comparable in order of magnitude
to those of the dysprosium condensate. Therefore, the creation of
soliton molecules in dipolar BECs appears achievable with current
experimental technologies.

Immersing an impurity atom into a Bose-Einstein condensate is an
advanced experimental procedure typically performed using techniques
like laser cooling and magnetic/optical trapping, followed by
precise control over the impurity's position and interaction
strength \cite{mayer2019}. We analyze the tunneling of an impurity
atom through a barrier in a double-well potential using the
two-state approximation given by Eq. (\ref{ansatz1}). If the
impurity is initially placed in a left well, represented as
$\phi(x,0) = \phi_L(x)$, the probability density $|\phi(x,t)|^2$
will oscillate periodically over time, with a period given by $T =
2\pi \hbar/\Delta$. Here, $\Delta = \mu_1 - \mu_0$ denotes the
energy difference between the first excited state and the ground
state. This difference is typically interpreted as the energy
splitting of the ground-state level in a single well, which occurs
due to the coupling between the two wells. In our calculations, the
length scale for the impurity atom is considered to be $\sim a_s$,
which yields the characteristic frequency $\omega_0 = \hbar/m_{Rb}
a_s^2 \simeq 26.5$ MHz for $m_{Rb}=1.4 \times 10^{-25}$~kg.
Therefore the qubit frequency shown in Fig. \ref{fig3}c is $\omega =
(2\pi/T_{meas}) \cdot \omega_0 \simeq 750$ kHz.

In this study, we considered the functioning of the proposed qubit
at the proof-of-principle level. Further research will focus on
exploring the influence of adverse effects, such as thermal noise,
phonon scattering, fluctuations in the soliton molecule, etc.

\section{Conclusions}

We demonstrated that a single impurity atom embedded in a dipolar
two-soliton molecule behaves as a controllable two-level system
undergoing coherent Josephson type oscillations which are
fundamental for the realization of qubits.  We employed a
variational approach to generate a two-soliton molecular potential,
allowing practical control over qubit characteristics, including
barrier height and tunneling frequency. The system is shown to be a
good candidate for a qubit with two-level system's purity ${\rm
Tr}(\rho^2) > 0.9999$, as shown in Fig.~\ref{fig5}. Future research
should explore entanglement effects and potential decoherence
mechanisms. The experimental realization of such a qubit using
chromium, dysprosium or erbium dipolar BEC systems appears feasible
within current ultracold-atom technologies.

\section*{Acknowledgements}

We gratefully acknowledge the support provided by King Fahd
University of Petroleum and Minerals (KFUPM) and the Research,
Development and Innovation Authority (RDIA) for funding this work
through project No. 22715-KFUPM-2023-KFUPM-B-4-1-EI.

\end{document}